\documentclass[%
 reprint,
superscriptaddress,
 amsmath,amssymb,
 aps,
 prx,
]{revtex4-2}

\usepackage{graphicx}
\usepackage{bm}
\usepackage{siunitx}
\usepackage[caption=false]{subfig}
\usepackage{hyperref}

\begin{document}


\title{An Integrated Photon-Pair Source with Monolithic Piezoelectric Frequency Tunability}

\author{T.~Brydges}
\email{tiffany.brydges@unige.ch}
\affiliation{Department of Applied Physics, University of Geneva, Geneva, Switzerland}
\author{A.~S.~Raja}
\affiliation{Laboratory of Photonics and Quantum Measurements, Ecole Polytechnique F\'ed\'erale de Lausanne (EPFL), Lausanne, Switzerland}
\author{A.~Gelmini}
\affiliation{Department of Applied Physics, University of Geneva, Geneva, Switzerland}
\author{G.~Lihachev}
\affiliation{Laboratory of Photonics and Quantum Measurements, EPFL, Lausanne, Switzerland}
\author{A.~Petitjean}
\altaffiliation[Current address: ]{Universit\'e C\^ote d’Azur, Institut de Physique de Nice, Nice, France}
\affiliation{Department of Applied Physics, University of Geneva, Geneva, Switzerland}
\author{A.~Siddharth}
\affiliation{Laboratory of Photonics and Quantum Measurements, EPFL, Lausanne, Switzerland}
\author{H.~Tian}
\affiliation{OxideMEMS Lab, Purdue University, West Lafayette, IN. 47907, USA}
\author{R.~N.~Wang}
\affiliation{Laboratory of Photonics and Quantum Measurements, EPFL, Lausanne, Switzerland}
\author{S.~A.~Bhave}
\affiliation{OxideMEMS Lab, Purdue University, West Lafayette, IN. 47907, USA}
\author{H.~Zbinden}
\affiliation{Department of Applied Physics, University of Geneva, Geneva, Switzerland}
\author{T.~J.~Kippenberg}
\affiliation{Laboratory of Photonics and Quantum Measurements, EPFL, Lausanne, Switzerland}
\author{R.~Thew}
\affiliation{Department of Applied Physics, University of Geneva, Geneva, Switzerland}

\date{\today}

\begin{abstract}
This work demonstrates the capabilities of an entangled photon-pair source at telecom wavelengths, based on a photonic integrated Si$_3$N$_4$ microresonator with monolithically integrated piezoelectric frequency tuning. Previously, frequency tuning of photon-pairs generated by microresonators has only been demonstrated using thermal control, however these have limited actuation bandwidth, and are not compatible with cryogenic environments. Here, the frequency-tunable photon-pair generation capabilities of a  Si$_3$N$_4$ microresonator with a monolithically integrated aluminium nitride layer are shown. Fast-frequency locking of the microresonator to an external laser is demonstrated, with a resulting locking bandwidth orders of magnitude larger than reported previously using thermal locking. These abilities will have direct application in future schemes which interface such sources with quantum memories based on e.g. trapped-ion or rare-earth ion schemes.
\end{abstract}

\maketitle


\section{\label{sec:level1}Introduction}

Quantum photonic sources and interfaces are predicted to be a crucial component of future quantum networks, enabling the transfer of information with unprecedented security compared to current classical protocols \cite{Thew:2007,Krenn:2016,Sangouard:2011,Wehner:2018}. In addition, the rapid development of quantum sensing, computation, and simulation technologies could eventually exploit reliable and robust quantum networks that distribute quantum information and entanglement as a resource. Such a network should also allow interfacing between the different technologies and platforms which may be used for these systems, such as photons \cite{Slussarenko:2019,Zhong:2020}, rare-earth ions \cite{Simon:2010,Businger:2020,Askarani:2021}, trapped ions \cite{Bruzewicz:2019, Brydges:2019,Krutyanskiy:20221}, and nitrogen-vacancy centres \cite{Bernien:2013,Pompili:2021}.\\

\noindent Integrated photonics is a promising solution to this problem. It allows large numbers of components to be packaged together in a compact and stable manner \cite{Moody:2022}, analogous to the packaging of electrical components in classical computers. Photonic integrated microresonators in particular have shown increasing potential for a wide range of applications over the past few years, including in coherent optical communication \cite{Palomo:2017}, frequency conversion \cite{Singh:2019}, and chip-scale frequency comb sources \cite{Kippenberg:2018}. As photon sources, microresonators have already demonstrated excellent performance \cite{Engin:2013,Silverstone:2015,Reimer:2016,Guo:2017,Faruque:2018,Samara:2019,Ma:2020,Oser:2020,Samara:2021,Samara:20211,Steiner:2021}. They are compact, highly stable and their heralding efficiencies have been shown to be comparable to some bulk optic, narrow-band cavity sources \cite{Rivera:2021,Samara:20211} -- a crucial characteristic for scaling quantum communication to larger numbers of photons \cite{Wang:2016,Chen:2017,Meyer:2017}. Silicon nitride ($\text{Si}_{3}\text{N}_{4}$) is an attractive platform for this technology, with low linear loss \cite{Blumenthal:2018,Liu:2021}, negligible two-photon absorption in the telecom band \cite{Rahim:2017}, and a wide transparency range -- from visible to mid-infrared wavelengths \cite{Munoz:2017}. The latter is especially advantageous with regards to interfacing between different quantum technology platforms, as some platforms are not at telecom wavelengths \cite{Simon:2010,Businger:2020,Askarani:2021}. Moreover, key components such as narrow-linewidth lasers \cite{lihachev2022low}, amplifiers \cite{Yang:2022}, and filters \cite{kumar2020integrated} can be fully integrated on a single chip, leading to mass manufacturable miniaturized quantum systems.\\

\noindent Integrated photonic $\text{Si}_{3}\text{N}_{4}$ microresonators have already shown they are well suited to the production of narrow-band photons at telecom wavelengths. More specifically, they have demonstrated bandwidths compatible with some rare-earth ion quantum memories \cite{Samara:2019,Samara:20211}, which are a cornerstone of many quantum repeater architectures \cite{Sangouard:2011, Rivera:2021}. However, interfacing a photon source with a quantum memory for such a quantum repeater-like architecture places very strict requirements on both the characteristics of the photons and the stability of the source itself. Of most relevance for this work is the ability for fast frequency tuning in order to achieve precise and stable frequency locking to the quantum memory. Typically, the ability to tune the resonance frequency of an integrated microresonator cavity (and by extension, the frequency of the generated photons) has only been possible by tuning the temperature of the device \cite{Engin:2013,Silverstone:2015,Faruque:2018,Samara:2019,Samara:2021,Samara:20211}. Temperature tuning is not ideal as it is relatively slow in comparison to electronic switching, with maximum actuation bandwidths on the order of \SI{10}{kHz} \cite{Xue:2016,Joshi:2016}, and impractical if the chip is to be used in cryogenic environments. In addition, temperature tuning is often monodirectional (i.e. active heating but passive cooling) \cite{Liu:2020}.

\begin{figure}[t]
\centering
\includegraphics[width=1\linewidth]{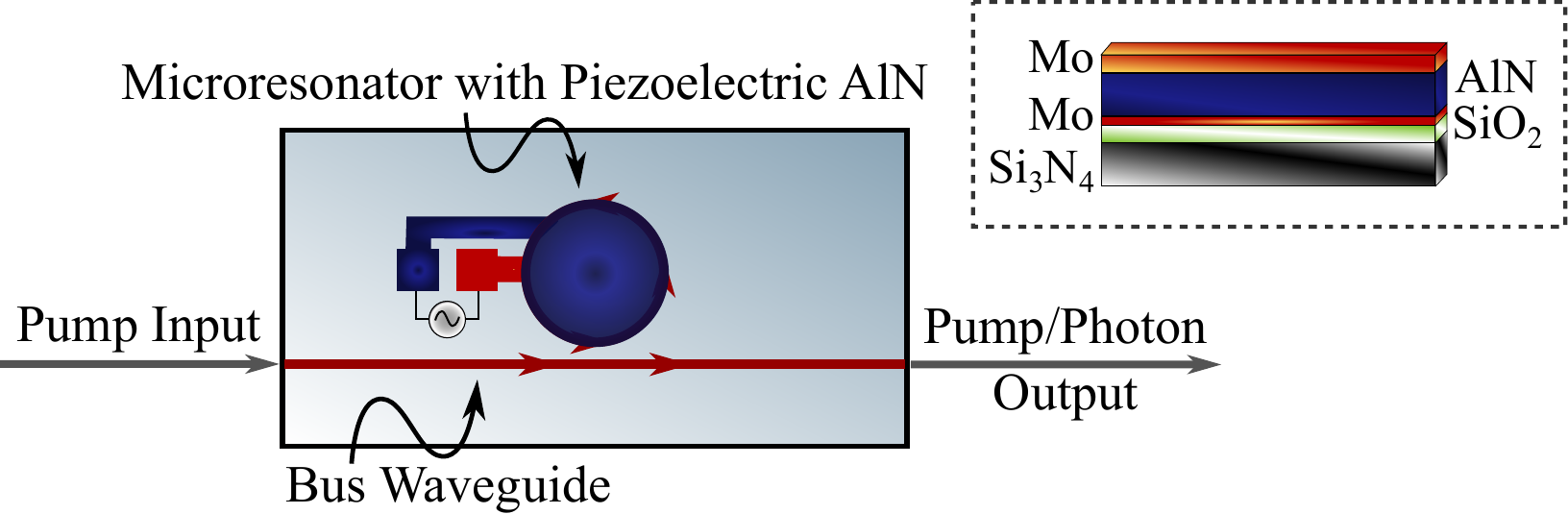}
\caption{Schematic of the electrically tunable microresonator. A monolithically integrated layer of AlN (dark blue), is placed on top of the existing $\text{Si}_{3}\text{N}_{4}$ structures. A voltage source is connected to the AlN layer in order to provide piezoelectric frequency tuning. \textbf{Inset:} Illustration of chip composition. The $\text{Si}_{3}\text{N}_{4}$ waveguide (black) is topped with a layer of silica cladding (green) to preserve the low propagation losses through the waveguide. A layer of molybdenum (red) acts as the bottom electrode, the AlN (dark blue) is the main piezoelectric material, with a final layer of molybdenum acting as the top electrode. See \cite{Tian:2020} for a detailed description of the chip design and fabrication process.}
\label{fig:TunRing}
\end{figure}

\noindent In contrast, integrated piezoelectric controllers have the potential to provide fast frequency tuning, which is both bidirectional as well as compatible with cryogenic environments \cite{Dong:2022}. Electronic tunability can be introduced to existing $\text{Si}_{3}\text{N}_{4}$ microresonator designs by the monolithic integration of an aluminium-nitride (AlN) piezoelectric actuator onto the $\text{Si}_{3}\text{N}_{4}$ microresonator \cite{Tian:2020,Liu:2020}, as illustrated in the inset of Figure \ref{fig:TunRing}. The application of a voltage to the actuator induces a mechanical stress, which alters the refractive index of the waveguide through the stress-optical effect \cite{Slot:2019}, so inducing a shift in the resonance frequency of the cavity. Such a design has already been successfully implemented in the classical regime through demonstrations of voltage controlled soliton initiation and stabilization \cite{Liu:2020}, unidirectional flow of light (isolators) \cite{Tian:2021}, and frequency-agile narrow-linewidth lasers \cite{lihachev2022low}. There have also been further experiments with this type of technology, such as the recently implemented high-speed programmable Mach-Zehnder meshes using AlN piezo-electric actuators coupled to $\text{Si}_{3}\text{N}_{4}$ waveguides \cite{Dong:2022}. However, up until this work, electronic tuning of photon frequencies generated by such microresonators, and the effect of the frequency tuning on the photon characteristics, has still not been investigated.

\begin{figure*}[t]
\centering
\includegraphics[width=1\linewidth]{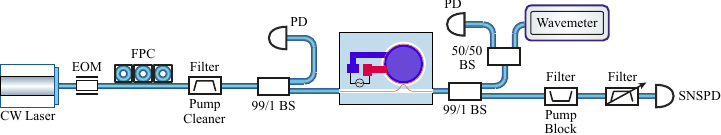}
\caption{Illustration of the basic experimental setup. The first filter before the microresonator is a dense wavelength division multiplexer (DWDM) centered at \SI{1555.75}{nm} (channel 27, \SI{200}{GHz} bandwidth) to provide a clean pump spectrum. The chip is mounted on a peltier which provides active temperature stabilization to within $\pm$\SI{10}{mK}. Ultra-high numerical aperture (UHNA) fibres are pigtailed to the input and output of the bus waveguide on the chip. The pump block filter after the microresonator is comprised of a set of 5 DWDMs to suppress the residual pump, and the tunable filter has a bandwidth of \SI{70}{pm}. Photons are detected using in-house developed SNSPDs \cite{Caloz:2018,Autebert:2020}. Photodiodes provide both power monitoring and a locking signal for the frequency stabilization. EOM: Electro-Optic Modulator, FPC: Fibre Polarization Controller, BS: Beam Splitter PD: Photodiode, SNSPD: Superconducting Nanowire Single Photon Detector.}
\label{fig:Setup}
\end{figure*}

\noindent This paper reports on a piezoelectric-controlled frequency-tunable source of entangled photon pairs, based on $\text{Si}_{3}\text{N}_{4}$ microresonators. The results show these microresonators have the ability for fast electronic frequency tuning of entangled photon-pairs, as well as high bandwidth locking to external lasers. This will be invaluable in the move towards interfacing with quantum memory platforms in the near future.

\section{\label{sec:level2}Materials \& Methods}

Figure \ref{fig:TunRing} illustrates the tunable photon-pair source. The microresonators are based on $\text{Si}_{3}\text{N}_{4}$ (shown as the black layer in the inset of Figure \ref{fig:TunRing}) with a silica cladding (green), and are fabricated using the photonic Damascene process \cite{Pfeiffer:2016}. The cladding layer is made to be sufficiently thick to ensure that the optical field is negligible at the top of the cladding layer, so preserving the low optical propagation loss of the $\text{Si}_{3}\text{N}_{4}$ waveguide \cite{Pfeiffer:2016}. A layer of molybdenum (red) is deposited on the top silica cladding to act as the ground electrode, the AlN piezoelectric actuator (dark blue) is the main piezoelectric material (thickness 1$\,\mu$m), with a final layer of molybdenum added to act as the top electrode. A detailed description of the chip design and fabrication process is given in \cite{Tian:2020}. The piezoelectric control is induced through the quasi-static stress-optic effect, causing strain in the amorphous $\text{Si}_{3}\text{N}_{4}$ waveguide and inducing changes in the refractive index of the $\text{Si}_{3}\text{N}_{4}$ \cite{Slot:2019}. This stress can be compressive or tensile, leading to either an increase or decrease in the local refractive index, and consequently an increase/decrease in the resonance frequency of the microresonator.\\

\noindent To demonstrate the frequency tuning capabilities of these devices, an initial characterisation was implemented using the scheme illustrated in Figure \ref{fig:Setup}. A CW laser at \SI{1555.7}{nm} passes through an electro-optic phase modulator (EOM) used to generate a Pound-Drever-Hall (PDH) error signal \cite{Drever:1983} for stabilization of the laser frequency to the microresonator. The light then passes through a polarization control stage to align the polarization with the quasi-transverse electric (TE) mode of the microresonator, and filtering stages to ensure a clean pump spectrum reaches the chip. Following this, the 1\% line from a 99:1 beam-splitter goes to a photodiode which is used to monitor the input power. The remaining 99\% is coupled directly into the bus waveguide on the chip through an ultra-high numerical aperture (UHNA) fibre pigtailed to the chip. When the laser is brought on resonance, the laser couples evanescently into the microresonator via the bus waveguide, generating signal and idler photons through spontaneous four wave mixing (SFWM) at each mode supported by the cavity. This leads to the generation of a photon frequency comb with a frequency spacing $\sim$\SI{200}{GHz}, corresponding to the free spectral range (FSR) of the microresonator. The device itself is an overcoupled device with an average intrinsic linewidth $\sim$\SI{40}{MHz} and total linewidth $\sim$\SI{200}{MHz}. The chip temperature is stabilized at \SI{297.32}{K} to within $\pm$\SI{10}{mK}, using an active PID stabilization system, which feeds back to a peltier on which the chip is mounted. The photons couple out of the ring back to the bus waveguide and out of the chip, again through a UHNA fibre pigtailed to the chip. The 1\% line of a second 99:1 beam-splitter goes to a second photodiode for power monitoring. Five dense wavelength division multiplexers (DWDMs) -- FSR~=~\SI{200}{GHz}, channel 27 -- suppress the residual pump light, with the resulting signal and idler photons passing through a tunable narrow-band filter (bandwidth of \SI{70}{pm}), being detected with in-house developed superconducting nanowire single photon detectors (SNSPDs) \cite{Caloz:2018,Autebert:2020}. By simultaneously changing the frequency of the tunable filter and each time acquiring photon counts, a spectrum of the photon frequency comb can be acquired. A measurement of this frequency comb is shown in Figure \ref{fig:appliedvoltages} (top). Further quantities characterising the source are given in the Supplementary Material.

\section{Results}

\begin{figure}
\centering
\includegraphics[width=.99\linewidth]{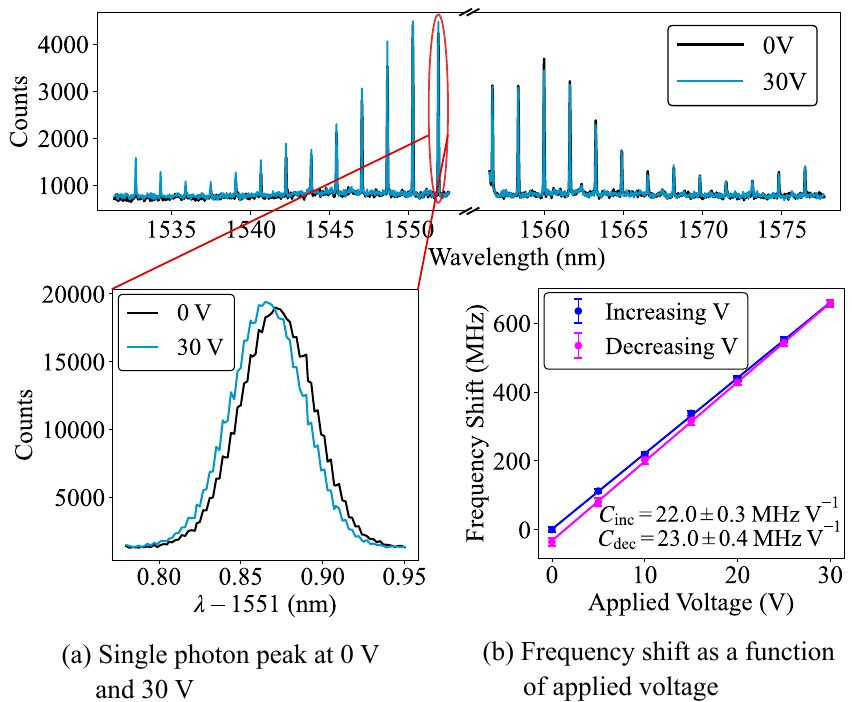}


\caption{\textbf{Top} Correlated photon pairs generated by the Si$_3$N$_4$ microresonator for two different applied voltages, \SI{0}{V} (black) and \SI{30}{V} (blue). The photon pairs form a frequency comb structure corresponding to the resonances of the cavity around the pump frequency. \textbf{a)} Measurement of the wavelength $\lambda$ of a single-photon peak (closest signal photon to the pump, red ellipse) for applied voltages of \SI{0}{V} and \SI{30}{V}. \textbf{b)} Graph of the measured single-photon frequency shift as a function of applied voltage. A slight hysteresis is observed between the measured shift when increasing the voltage (blue), and when decreasing the voltage (pink). The extracted tuning coefficients from weighted fits are $C_{\text{inc}} = 22.0\pm$\SI{0.3}{MHz} $\text{V}^{-1}$ for the increasing voltage, and $C_{\text{dec}} =23.0\pm$\SI{0.4}{MHz} $\text{V}^{-1}$ for the decreasing voltage. The shift was extracted through a Gaussian fit to the single-photon peak. Error bars are the standard error from the fit.}
\label{fig:appliedvoltages}
\end{figure}

By applying a voltage to the actuator, the frequencies of the modes supported by the microresonator (and hence the photon frequency comb) shift by a set amount. This shift was measured by scanning the tunable filter over a single-photon peak for different applied voltages. Figure \ref{fig:appliedvoltages} a) shows such single-photon peaks for two different applied voltages. It can be seen that the application of \SI{30}{V} induces a shift of approximately \SI{600}{MHz}. Figure \ref{fig:appliedvoltages} b) shows these frequency shifts as a function of the applied voltage, with an average linear tuning coefficient of \SI{22.5}{MHz}~$\text{V}^{-1}$. A single experiment cycle consisted of measuring the induced frequency shift for increasing voltages up to \SI{30}{V}, followed immediately by measuring the shift for decreasing voltages back to \SI{0}{V}. Small, linear drifts in the microresonator resonance frequency were observed due to changes in the ambient temperature over the long timescales of each experiment cycle. These were measured before and after each experiment cycle by sending part of the residual pump light after the microresonator to a wavemeter. The average of these drifts was consequently subtracted off the measurement cycle occurring in-between. A slight hysteresis is observed, consistent with that seen in \cite{Liu:2020}, which is most likely due to trapped charges \cite{Aardahl:1999}. This is, to the authors' knowledge, the first demonstration of piezoelectric-controlled frequency-tuning of photon pairs generated by microresonators.

\subsection{Bandwidth Invariance}

\begin{figure}
\centering
\includegraphics[width=.99\linewidth]{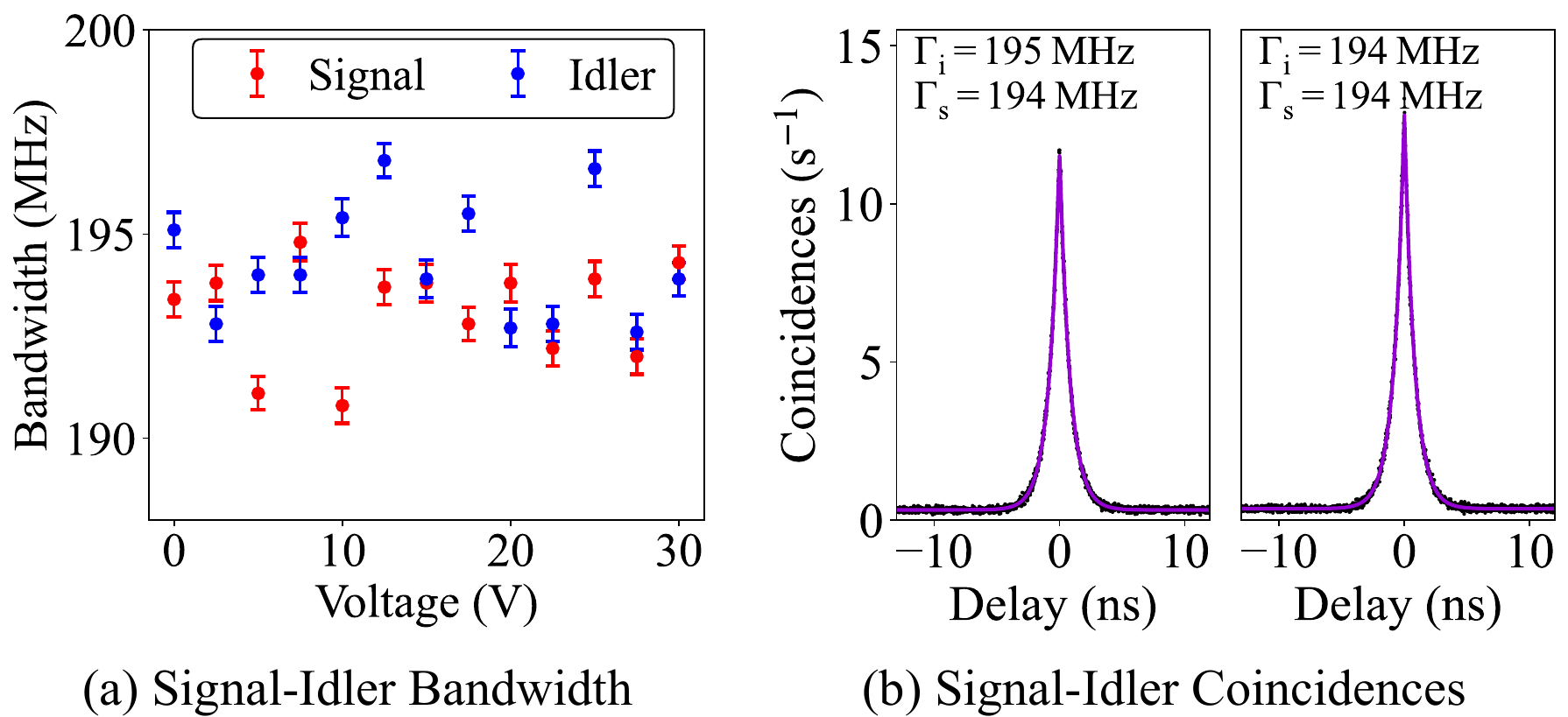}
\caption{\textbf{a)} Bandwidth of the signal (red) and idler (blue) photons as a function of applied voltage. The bandwidth is extracted directly from the linewidth of the signal-idler coincidence peak. Error bars are the standard error. \textbf{b)} Signal-idler coincidence peaks for applied voltages of \SI{0}{V} and \SI{30}{V}. The linewidth of both the signal and idler photons $\Gamma_{s,i}$ are $\sim$\SI{194}{MHz} for both voltages.}
\label{fig:bandwidths}
\end{figure}

Quantities such as the photon-bandwidth are dependent on the refractive index of the cavity material, with FWHM $\propto1/n_{\text{g}}$, where FWHM is the Full Width at Half Maximum of the cavity resonance, and $n_{\text{g}}$ the group refractive index \cite{Bogaerts:2012}. It is therefore natural to question whether there will be a significant change in the photon-bandwidth with the applied voltage (as the voltage induces changes in the refractive index of the cavity). This is of particular importance when interfacing with rare-earth ion quantum memories, where the photon bandwidths generated by the microresonator should match the absorption profile of the memory, and remain constant irrespective of any applied voltage. If the photon bandwidths change such that they increase above that of the memory, then the memory would begin to act as a filter, with consequently fewer photons stored \cite{Clausen:2014}. This would lead to a decrease in the so-called heralding efficiency, which is the efficiency with which the photon stored by the memory is `heralded' by detection of its partner photon \cite{Rivera:2021}. To experimentally measure if the photon bandwidths are voltage dependent, a 50/50 beam splitter was included after the pump suppression filter in the setup shown in Figure \ref{fig:Setup}, with each arm passing through a tunable filter set to either the signal or idler wavelength. Each arm then goes to its own SNSPD, with the coincidences between detections subsequently recorded. Examples of the signal-idler coincidences recorded from this setup are shown in Figure \ref{fig:bandwidths} b) for two different voltages applied to the AlN layer. The photon bandwidth can then be directly extracted as the linewidth of the signal-idler coincidence peak. The coincidence peak is fitted using a convolution of a double-exponential decay convolved with a Gaussian \cite{Samara:20211}. This takes into account the temporal shape of each photon (defined by the cavity), along with the temporal jitter introduced from the SNSPDs, which is a Gaussian distribution \cite{Caloz:2018}. Note that the detector jitter is much smaller than the bandwidth of the photons.\\

\noindent The measured bandwidth of the signal and idler photons as a function of the applied voltage is shown in Figure \ref{fig:bandwidths} a). It can be seen from the figure that there appears to be no measurable correlation between the bandwidth of the photons and the voltage applied to the microresonator. To confirm that the scatter seen in the figure is also not significant, the overlap between two wavepackets which are assumed to differ in bandwidth by the maximum scatter of \SI{10}{MHz} seen in Figure \ref{fig:bandwidths} a) can be calculated, and is found to be $\sim$97\%. From this, it can be concluded that there should be no effect on the heralding efficiency from application of voltages to the resonator.

\subsection{Frequency Stabilization of the Microresonator}

When interfacing these microresonator photon sources with quantum memories and moving towards a future quantum repeater-like implementation, the photon frequency must be resonant with the quantum memory at all times. As such, a frequency locking scheme will be needed to lock the resonance of the microresonator to the frequency of the quantum memory. There have been previous demonstrations of frequency locking microresonators by feeding back a locking signal to the temperature of the microresonators, however the achievable locking bandwidth in such schemes is very small, previously reported to be only in the tens of Hertz regime \cite{Padmaraju:2014,Zhu:2014,Zektzer:2016,Zhu:2019} (although the actuation bandwidth is much higher \cite{Xue:2016,Joshi:2016}). Focus has therefore moved away from such thermal locking schemes, with recent developments including piezoelectric frequency-locking schemes integrated on-chip \cite{Wang:2022}. However, the use of such devices as photon-pair sources has not yet been demonstrated.\\

\begin{figure}
\includegraphics[width=.99\linewidth]{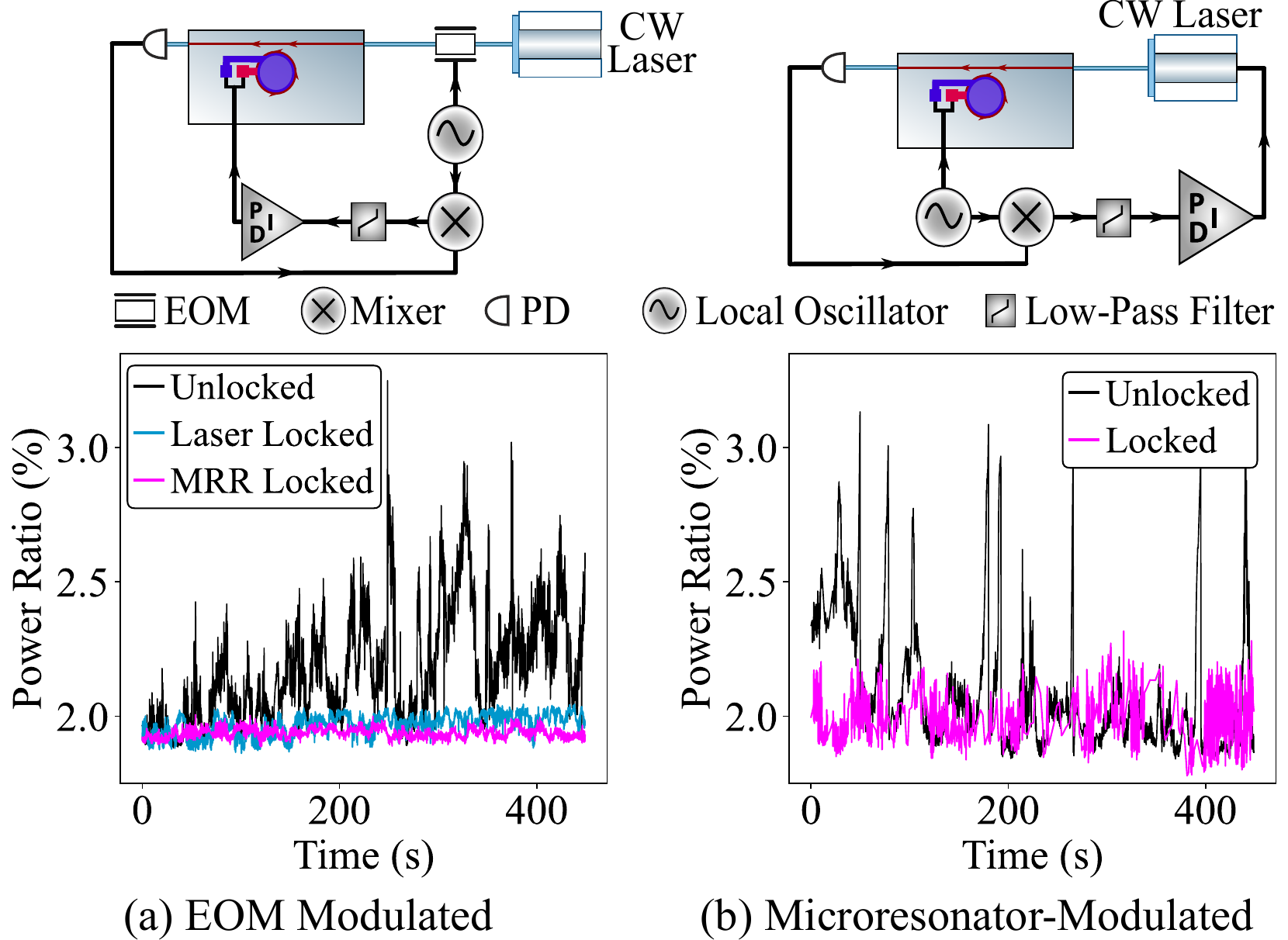}
\caption{Comparison of transmitted power ratio for unlocked and locked circuits. Above each graph, an illustration of the relevant locking circuit is shown. \textbf{a)} For modulation of the EOM, a comparison of the unlocked (black) and locked (magenta) cases is shown when the locking signal is fed back to the chip as shown in the illustration. The power ratio when the locking signal is fed back to the laser (blue) is also shown for comparison. \textbf{b} For microresonator-modulation, a comparison of the unlocked (black) and locked (magenta) cases is shown, with the locking signal subsequently fed back to the laser.}
\label{fig:powerratiosnorm}
\end{figure}

\noindent With the AlN integrated microresonator devices, it is possible to lock the cavity resonance to the pump laser using the PDH technique \cite{Drever:1983} and feeding back the signal to the AlN layer \cite{Liu:2020}. A schematic of this scheme is shown in Figure \ref{fig:powerratiosnorm} a). An error signal was generated by sending a local oscillator modulation signal to an external EOM and monitoring the residual pump light transmitted by the microresonator when on resonance with the laser, using a photodiode. The resultant optical modulation will either be in-phase or out-of-phase with the applied local oscillator modulation, depending on whether the laser frequency is positively or negatively detuned from the resonance frequency. By mixing the resulting optical modulation signal with the applied local oscillator, and passing the signal through a low-pass filter in order to suppress higher-order frequency components, an asymmetric error signal is obtained, which can be passed to a PID servo loop. The resultant voltage from the PID servo loop was subsequently applied to the AlN layer on top of the microresonator cavity, so adjusting the frequency of the cavity such that it remained resonant with the pump laser. This scheme mimics that which will be needed to keep the microresonator resonant with a quantum memory in the future.\\

\noindent Figure \ref{fig:powerratiosnorm} a) shows the results from this locking scheme. The plot shows the power ratio, which is the ratio of the transmitted pump power after the microresonator to the input power of the microresonator. The power ratio provides an indication of the stability of the laser with respect to the microresonator resonance as, at the point of resonance, it will be at a minimum. Deviations away from this point cause the power ratio to increase. Shown in black is the power ratio for the unlocked resonance, where large fluctuations can be seen as the laser drifts in and out of resonance with the microresonator. The locked case is shown in magenta, where the power ratio is extremely stable in comparison. For further comparison, the locked case when the locking signal is instead fed back to the laser (rather than the AlN layer) is shown in blue. The implementation of this system allowed the chip to remain on resonance with the laser for in excess of 16 hours. The system eventually fell out of lock when the locking circuit reached its maximum value of \SI{30}{V} (that is, to keep the chip on resonance with the laser, the circuit would have had to provide more than \SI{30}{V}). Even though AlN itself can sustain in excess of $\pm$\SI{100}{V} \cite{Liu:2020}, in this case the maximum was restricted to \SI{30}{V} in order to avoid damage to the wirebonds attached to the chip. This experiment demonstrates that we have a source of photon-pairs which can be frequency stabilized to an external reference laser, compatible with the requirements for some quantum memory schemes \cite{Businger:2020}.

\subsection{Microresonator-Modulated Locking}

The technique of modulating an external EOM to implement a PDH lock has been widely implemented for many decades, however there also exist extensions to the original proposal where the modulation signal is applied to different (potentially integrated) components. For example, Padmaraju et al. \cite{Padmaraju:2014, Zhu:2014} have previously implemented a scheme whereby the modulation signal is applied to the microresonator itself via a resistive heater. There are limitations to the locking bandwidth of this implementation, however, primarily due to the slow thermal response of the chip \cite{Padmaraju:2014, Zhu:2014}. With the microresonators presented here, a modulation signal can instead be applied to the piezoelectric layer, as demonstrated in similar scenarios in \cite{Liu:2020,Wang:2022}. This leads to a modulation of the optical signal, and the derivation of a PDH error signal which can be used to lock the pump laser to the microresonator, as shown in the schematic in Figure \ref{fig:powerratiosnorm} b). The plot in Figure \ref{fig:powerratiosnorm} b) shows power ratio measurements for the unlocked case where the microresonator was modulated at \SI{40}{MHz}, and for the locked case, again with the microresonator modulated at \SI{40}{MHz}. Although noisier than when modulating the external EOM, it can clearly be seen that the laser is kept on resonance with the chip throughout the measurement, in contrast to the unlocked case where it drifts in and out. The increased noise is likely due to the sub-optimal locking regime which had to be used to implement this scheme -- see the Supplementary Material for further information. This novel demonstration of microresonator-modulated fast-frequency locking shows the potential use of this technology in the future for removing the need for often-expensive EOMs, which are also not necessarily available at all required wavelengths.

\subsection{Locking Bandwidth}

\begin{figure}[t]
\centering
\includegraphics[width=.99\linewidth]{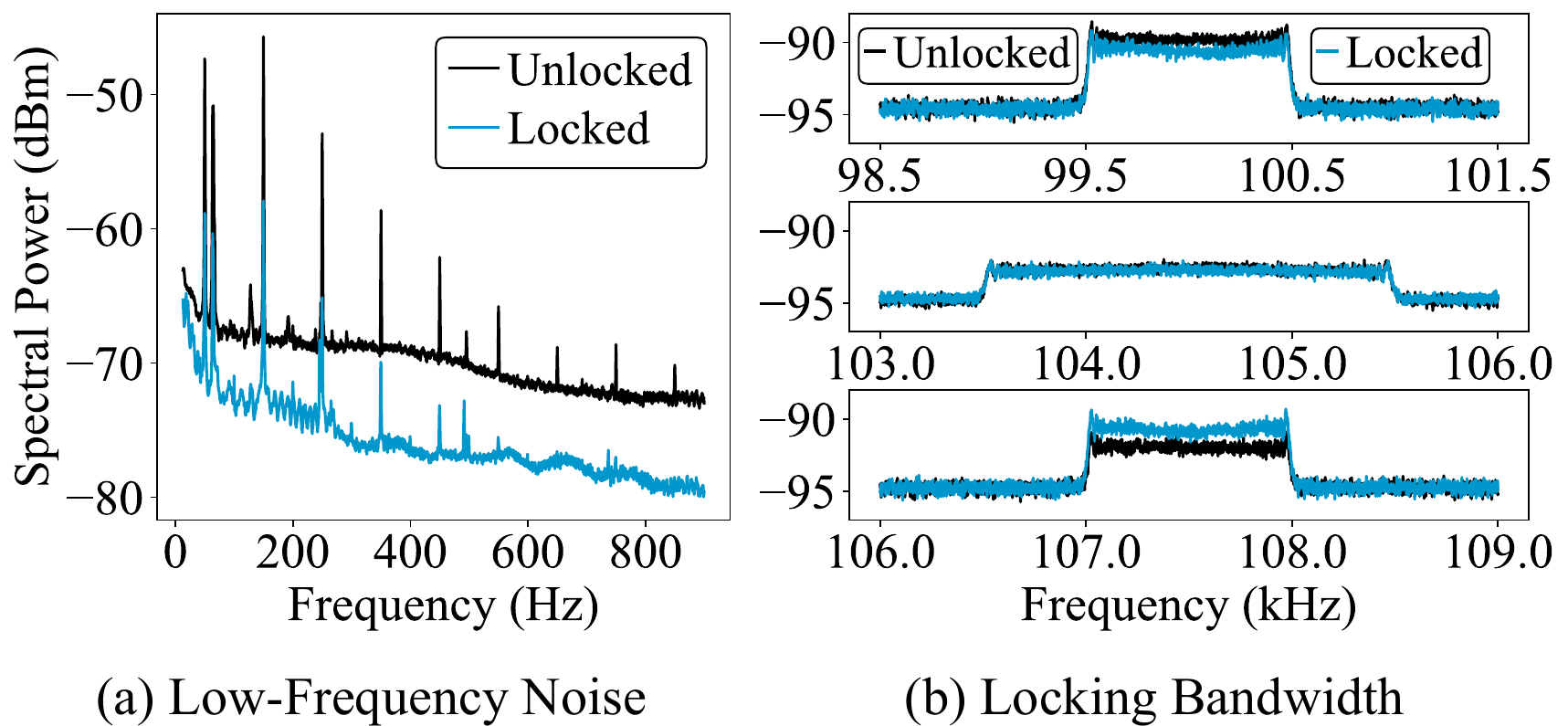}
\caption{Measured spectral power after the microresonator for both unlocked (black) and locked (blue) circuits. \textbf{a)} Measured spectral power for low frequencies. \textbf{b)} Determination of locking bandwidth using the measured spectral power for several higher frequencies. Shown is the point below crossing (top), at the point of crossing (middle), and above crossing (bottom). See text for further details. A Gaussian filter was applied to the data to better resolve the characteristics.}
\label{fig:noisecomparison}
\end{figure}

Finally, to measure the response of the microresonator-locking system (schematic in Figure \ref{fig:powerratiosnorm} a)) to noise, an external noise process was introduced to the system by applying a linear frequency sweep to the laser current, with part of the output from the microresonator monitored by a photodiode. Figure \ref{fig:noisecomparison} a) shows the measured spectral power from the FFT of the photodiode signal for the unlocked (black) and locked (blue) systems with a noise sweep from \SI{1}{Hz} to \SI{1}{kHz} (a Gaussian filter was applied to the data to better resolve the characteristics). Overall, the intensity noise is suppressed by a factor of $\sim$\SI{8}{dB} in this region. However, there are also several noise spikes which are suppressed by a factor well in excess of this. By driving the system with increasingly higher noise frequencies, this allowed an estimation of the bandwidth of the entire locking circuit (an amalgamation of the response of the microresonator AlN layer and the electronic circuitry) to be made. By looking for the point at which the unlocked and locked spectral powers overlap, so the noise bandwidth that the entire circuit can suppress can be estimated. Figure \ref{fig:noisecomparison} b) shows three graphs around this point of interest: The upper plot shows the noise regime from \SI{99.5}{kHz} to \SI{100.5}{kHz}, where the locked circuit still gives slight suppression of the noise compared to the unlocked circuit. The middle plot shows the regime from \SI{103.5}{kHz} to \SI{105.5}{kHz}, where the two circuits are the same. The bottom plot shows the regime from \SI{107}{kHz} to \SI{108}{kHz}, where the locked case actually has a higher noise level than the unlocked case. From this, the locking bandwidth can be estimated to be on the order of \SI{100}{kHz}. This is a vast improvement on those schemes which use the thermal response of the chip for stabilization, with previous reports only reaching bandwidths on the order of tens of Hertz \cite{Padmaraju:2014,Zhu:2014,Zektzer:2016}.

\section{Conclusion}

This work has demonstrated the suitability of these $\text{Si}_{3}\text{N}_{4}$ based microresonators, with monolithically integrated piezoelectric actuators, as frequency-tunable photon-pair sources. It has shown that they continue to produce narrow-linewidth photon-pairs comparable to previous state-of-the-art sources which lack such piezoelectric-controlled frequency tunability. They provide reliable, voltage-controlled frequency tuning of the generated photon pairs, as well as allowing for PDH locking with bandwidths far in excess of those achievable using temperature locking. These abilities add to the existing `photonic toolbox' available when using such integrated microresonators as photon-pair sources, and will have direct application when interfacing with quantum memories based on trapped-ion or rare-earth ion schemes in the near future.

\begin{acknowledgments}
The authors would like to thank L. Stasi and G. V. Resta for development, maintenance, and technical support of the SNSPDs and cryostats, and C. Barreiro for electronic support. The authors would also like to thank  A. Toros, V. Shadymov and A. Voloshin for assistance in wirebonding. T.B. would like to thank J. A. Harrington for useful discussions. The chip samples were fabricated in the EPFL centre of MicroNanoTechnology (CMi), and in the Birck Nanotechnology Center at Purdue University. AlN deposition was performed at OEM Group Inc.\\
This work was supported by: The Swiss National Science Foundation SNSF (Grant No. 200020$\_$182664); the Swiss National Science Foundation SNSF (Grant No. 192293); the NCCR QSIT; Contract W911NF2120248 (NINJA) from the Defense Advanced Research Projects Agency (DARPA), Microsystems Technology Office (MTO). A.S. acknowledges support from the European Space Technology Centre with ESA Contract No. 4000135357/21/NL/GLC/my
\end{acknowledgments}

\appendix

\section{Characteristics of the Microresonator Photon-Pair Source}

\vspace{2mm}
This section presents characteristics of the voltage-controlled, frequency-tunable microresonator photon-pair source.\\

\noindent The Si$_3$N$_4$ microresonator has waveguides of dimension 2.2$\,\mu$m width and \SI{900}{nm} height, with an AlN layer of thickness 1$\,\mu$m. It is an overcoupled device, with $\kappa_{\text{ex}} \approx 3 \times \kappa_{0}$, where $\kappa_{\text{ex}}$ is the external coupling rate from the bus waveguide to the cavity, and $\kappa_0$ corresponds to the intrinsic linewidth of the cavity. This leads to an improved extraction efficiency, with the device having an average intrinsic linewidth $\sim$\SI{40}{MHz} and total linewidth $\sim$\SI{200}{MHz}.\\

\begin{figure}
\centering
\includegraphics[width=.99\linewidth]{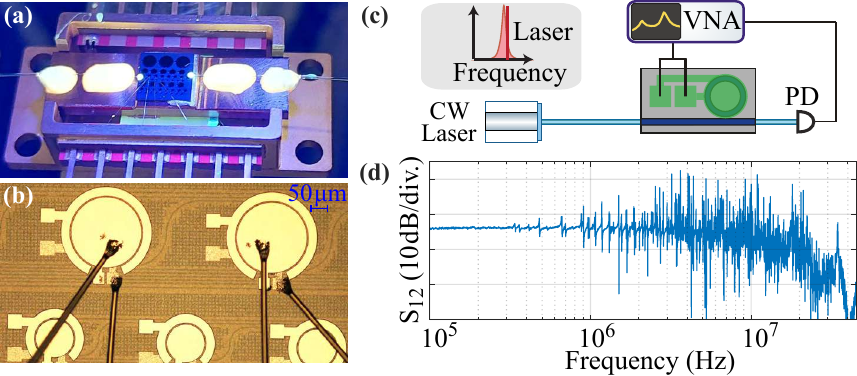}
\caption{a) Photograph of the electrically and optically packaged chip. The fibres are pigtailed to the chip using epoxy. b) Microscope image of the electrically wire bonded AlN actuator on top of the Si$_3$N$_4$ microresonator. c) \& d) Optomechanical frequency response $S_{21}(\omega)$ of the microresonator when modulated at frequency $\omega$. c) Experimental scheme used to measure the optomechanical frequency response. d) Experimental results from a microresonator similar to the device used throughout this work, except without optical packaging. VNA: Vector Network Analyser, PD: Photodiode.}
\label{fig:acres}
\end{figure}

\noindent Figure \ref{fig:acres} a) and b) shows a photograph of the electrical and optical packaging of the chip. The UHNA fibres are pigtailed to the chip using epoxy. The electrical wirebonds are made from 17$\,\mu$m diameter aluminium. Figure \ref{fig:acres} c) and d) shows the optomechanical frequency response $S_{21}(\omega)$ of a similar device when a modulation at frequency $\omega$ is applied to the AlN layer of the microresonator (i.e. a measure of the electrical to optical transduction). The experimental setup of the actuation response measurement is shown in Figure \ref{fig:acres} c). An actuation voltage at frequency $\omega$ is generated from a vector network analyser (VNA), and applied to the piezoactuator on the chip. The pump laser frequency is tuned to the slope of the Si$_3$N$_4$ microresonator resonance, with a photodiode detecting the optical signal containing the modulation. Figure \ref{fig:acres} d) shows the measured optomechanical $S_{21}(\omega)$ response of a tunable microresonator device similar to that used throughout this work, except without optical packaging. The actuation applied to the AlN piezoelectric layer excites many mechanical bulk or contour modes of the photonic chip, leading to a non-flat actuation response. It can be seen that the response is flat up to \SI{340}{kHz}, where the first mechanical resonance is visible. Note that these mechanical modes can be suppressed via techniques such as mode-cancellation, apodization, or by using an acoustic absorber such as carbon tape or a glass submount \cite{Siddharth:2021}.\\

\noindent The bandwidth of the photon pairs (shown in Figure 4 of the main text) gives an indication of the Q-factor of the microresonator. For the microresonator used in this work, the Q-factor was calculated to be $\sim$$1\times10^{6}$ (equivalent to an average photon bandwidth of \SI{194}{MHz}), comparable to many state-of-the art $\text{Si}_{3}\text{N}_{4}$ photon-pair sources \cite{Wu:2021}.\\

\noindent The intracavity pair generation rate (PGR), defined as:
\begin{equation}
\text{PGR}= \frac{S_{\text{i}}S_{\text{s}}}{R_{\text{cc}}},
\end{equation}

\noindent with $S_{\text{i(s)}}$ the single-photon count rates for the idler(signal) photons and $R_{\text{cc}}$ the coincidence rate, was found to be on the order of PGR = $32\times10^{3}\,\text{s}^{-1}\,\text{mW}^{-2}$. This is comparable to many other microresonator photon pair sources which lack the voltage-controllable frequency tunability of these devices, such as those in \cite{Samara:20211,Wen:2022}.
For the results shown in Figure 4 a) of the main text, with an average measured bandwidth of \SI{194}{MHz}, this results in a probability of pair generation per coherence time of $p=0.01$.\\


\noindent The heralding efficiency, defined by:
\begin{equation}
\eta_{\text{h},\text{i(s)}} = \frac{R_{\text{cc}}}{S_{\text{s(i)}}\eta_{\text{d}}},
\end{equation}

\noindent is found to be, on average, $\eta_{\text{h}}=1.6\%$ for both signal and idler, using a detector efficiency of $\eta_{\text{d}}=0.84$ for both detectors. As the aim of this work is to demonstrate the frequency-tuning capabilities of the photon-pair source, and not to optimise photon-pair production, a large number of DWDMs were used, as well as relatively high-loss tunable filters. Therefore, a more useful measure is the coupling efficiency from the chip into the UHNA fibre. The coupling efficiency is given by:
\begin{equation}
\eta_{\text{coup},\text{i(s)}} = \frac{R_{\text{cc}}}{S_{\text{s(i)}}\eta_{\text{d}} t_{\text{i(s)}}},
\end{equation}
where $t_{\text{s(i)}}$ is the overall transmission from the output of the UHNA fibre pigtailed to the chip to the detector for the signal(idler) path. The average coupling efficiency was found to be $\eta_{\text{coup}}=14.7\%$, for both signal and idler.\\

\begin{figure}[t]
\centering
\includegraphics[width=0.8\linewidth]{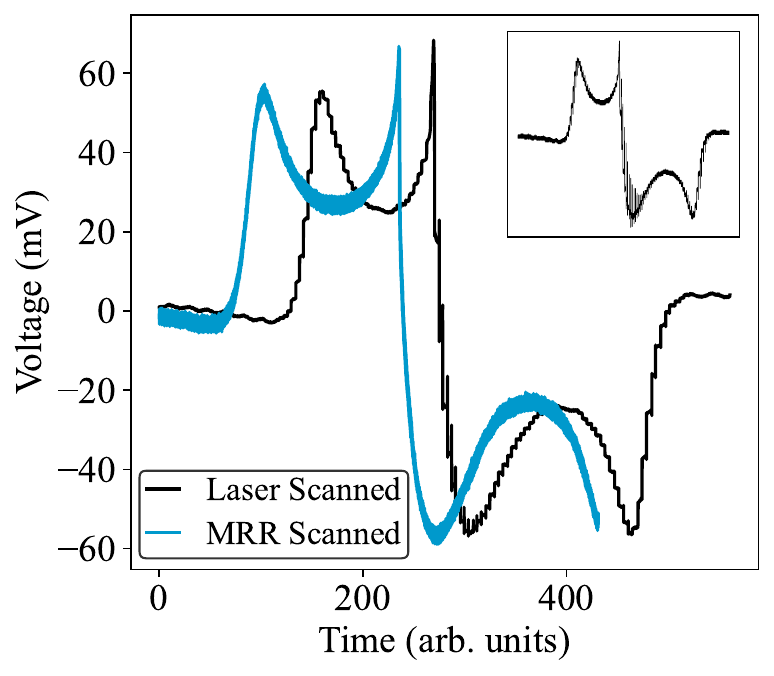}
\caption{a) Oscilloscope trace of the error signals generated when a local oscillator is applied to an external EOM. In black, the laser piezo was scanned over the microresonator point of resonance. In blue, the microresonator point of resonance was scanned over the laser frequency. \textbf{Inset:} Unsmoothed data. Data was smoothed by a polynomial of order 7 in order to remove noise present from the laser's piezo.}
\label{fig:ErrsigPieCav}
\end{figure}

\begin{figure}[t]
\centering
\includegraphics[width=0.99\linewidth]{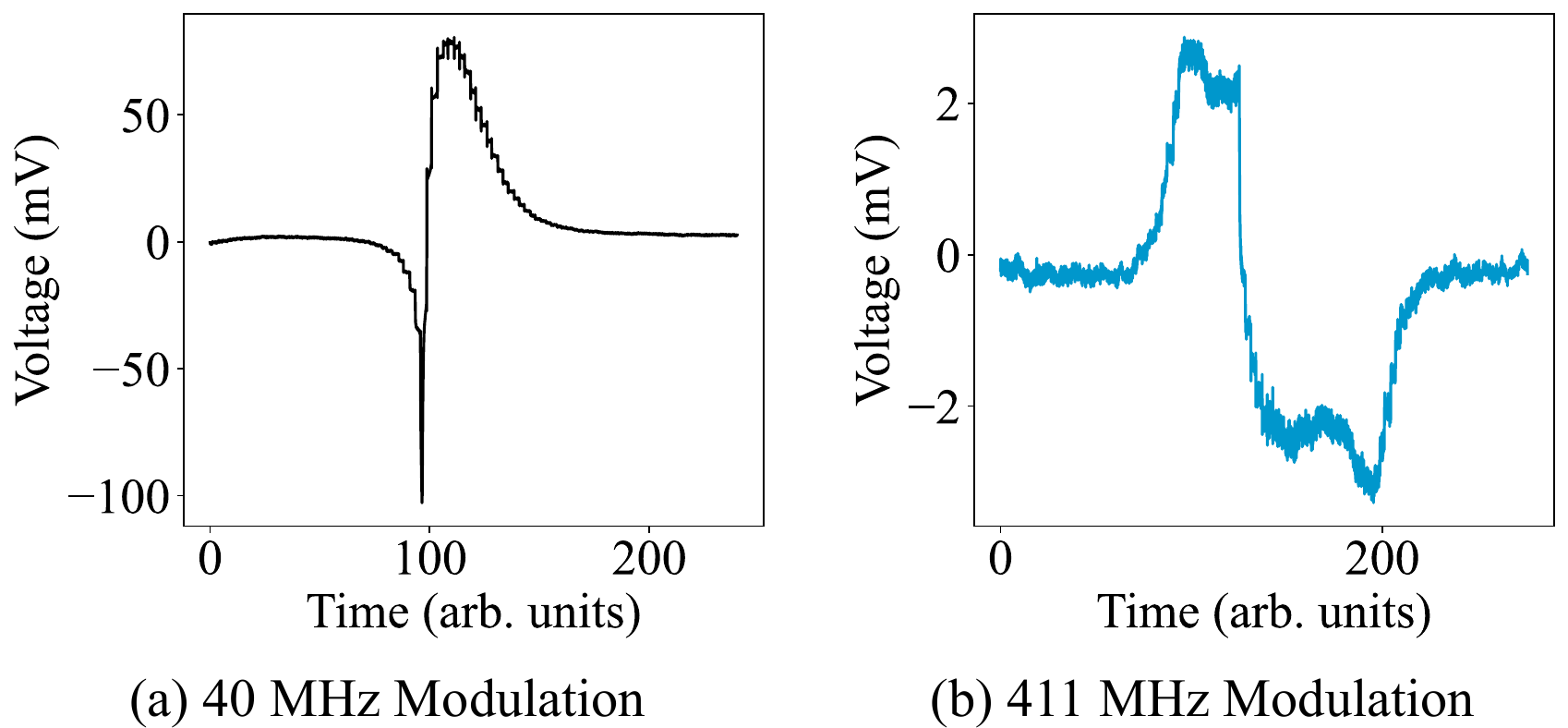}
\caption{Oscilloscope traces of error signals generated when applying a modulation signal to the microresonator itself, and scanning the laser piezo over the point of resonance. a) Error signal when the microresonator is modulated by a signal of \SI{40}{MHz}. b) Error signal when the microresonator is modulated by a signal of \SI{411}{MHz}. Note the difference in voltage scale between this plot and b). Data was smoothed by a polynomial of order 7 in order to remove noise present from the laser's piezo.}
\label{fig:chipmod}
\end{figure}

\noindent Improvements to the heralding efficiency could be straightforwardly implemented by replacing the tunable filters with fixed-frequency DWDMs at channels corresponding to appropriate signal/idler generation wavelengths. However, these were not used here in order to allow for scanning over the photon frequencies to measure the relevant shifts induced by voltages applied to the microresonator.

\section{Characteristics of the Locking Schemes}
\vspace{2mm}
Figure \ref{fig:ErrsigPieCav} shows the error signals generated when a modulation frequency of \SI{846.6}{MHz} is applied to an external EOM. Shown in black is the signal extracted when a laser piezo is scanned over the microresonator point of resonance. As the laser piezo introduced noise spikes to the error signal (inset), the data was smoothed using a 7th order polynomial to better see the signal characteristics. In blue is shown the error signal extracted when instead the microresonator is scanned, through application of a voltage to the AlN layer, over the laser frequency. The overall characteristics of the two error signals can be seen to be identical.\\

\noindent Figure \ref{fig:chipmod} b) and c) shows the error signals extracted by applying a modulation signal to the microresonator itself at two different frequencies, while simultaneously scanning the laser frequency over the point of resonance. In the ideal locking regime, the microresonator would be modulated at a frequency much larger than the resonance frequency (e.g. at a frequency similar to the \SI{846.6}{MHz} applied to the external EOM). However, it can be seen from Figure \ref{fig:ErrsigPieCav} c) that at higher modulation frequencies, the amplitude of the error signal is significantly reduced compared to lower frequencies. As such, a lower modulation frequency had to be used for locking of the laser to the microresonator for the microresonator-modulated case than with the EOM. It is likely that this sub-optimal locking regime contributes to the increased fluctuations present in the locked power ratio shown in Figure 5 b) of the main text. If a high-frequency modulation was chosen to coincide with a HBAR mode of the chip, it is likely that the amplitude of the error signal would be greatly enhanced, as shown previously in \cite{Liu:2020}.

%

\end{document}